# A Novel mapping for visual to auditory sensory substitution

Eysan Mehrbani
*Electrical Engineering Department,
Iran University of Science and Technology,*
Tehran, Iran
aysan_ghabous@elec.iust.ac.ir

Seyedeh Fatemeh Mirhosseini[1]
*Electrical Engineering Department,
Amirkabir University of Technology*
Tehran, Iran
f_mirhoseini@aut.ac.ir

Noushin Riahi
*Computer Engineering Department
Alzahra University*
Tehran, Iran
nriahi@alzahra.ac.ir

*Abstract*— visual information can be converted into audio stream via sensory substitution devices in order to give visually impaired people the chance of perception of their surrounding easily and simultaneous to performing everyday tasks. In this study, visual environmental features namely, coordinate, type of objects and their size are assigned to audio features related to music tones such as frequency, time duration and note permutations. Results demonstrated that this new method has more training time efficiency in comparison with our previous method named VBTones which sinusoidal tones were applied. Moreover, results in blind object recognition for real objects was achieved 88.05 on average.

*Keywords*— *sensory substitution, audio-vision, semantic segmentation, Mask-RCNN, note permutation.*

## I. Introduction

On the case of many visually impaired individuals, the functionality of the brain is intact while the vision sensory module is incapable of appreciating spatial information. Sensory substitution which relies on other sensory modalities attempts to transmit the surrounding information in an efficient approach in terms of comprehensibility, promptness, convenience, and being limited cognitive load.

Using the auditory and tactile sense as a means of transferring visual information were the most common propositions regarding the visually impaired [1]. Yet the primary attempts relied on the tactile sense such as Braille notations and Bach-y-Rita who developed systems capable of making blind individuals able to perceive visual information through Haptics [2]. Visual to tactile substitution are basically focused on providing users with distance measurements from nearby obstacles so as to give them a rough, yet intuitive understanding of their surroundings [1].

## II. Related Work

Starting with the primary work of the vOICe [3] visual to auditory sensory substitution was introduced. The vOICe device set is consisted of a camera capturing a gray scale image and a pair of earphones playing the corresponding sound with the current frame. The algorithm converting the frame to a short audio stream assigned a monotone sinusoid waveform to each pixel row. The angular frequency of the sinusoids are increased for higher rows in the image frame. The image is scanned column by column and rightwards with each column retaining a short sound as the summation of pixel sounds. The pixel sound is louder if the analogous intensity level is higher. The column sounds are concatenated to form a total stream [4]. Figure 1 illustrates the mapping proposed by Peter B. Meijer [3].

The vOICe achieved notorious results with long-term users enabling them to develop a perception towards concepts of edge, contrast, form, depth, and even color [5] even though the algorithm does not code the depth and color. Amir Amedi et. al. [6] in their algorithm the EyeMusic, have added the RGB color channels into the image and developed a coding scheme similar to the vOICe but with three sound channels each with a unique and distinguishable timbre. The selected instruments are generally able to produce a continuous pitch in order to form the long lasting notes. The same principal also hold in the present work as the length of the notes comprises information of object widths. J. Ward in [7] also tried coding the color into the vOICe main algorithm by using tactile patches accompanying the sound waves as color information enhances object-ground segmentation, and provides more stable cues for scene and object recognition. In [8] the auditory substitution of visual sense is claimed superior to the tactile in terms of filling the modality gap.

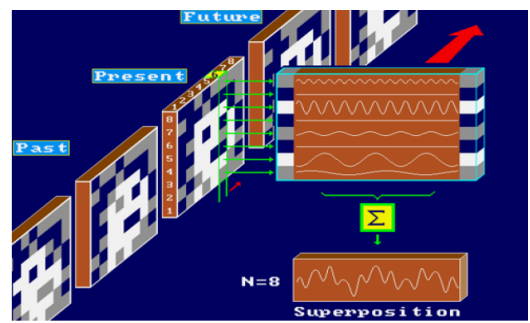

Fig. 1. Image to audio conversion using the vOICe algorithm.

There are other novel approaches using various sensors in helping the blind navigate rather than simply perceive the environment. For instance, the Smart Cam System [9] used a depth camera and a server for SLAM processing that allowed for six degrees-of-freedom indoor location, plus obstacle detection features. Using compatible hardware platforms and haptic actuators embedded in a cane [10] exploits a wayfinding and obstacle detection method.

There is an ongoing trend towards using smartphones in sensory substitutions as a portable and accessible device for everyone. The compatibility of smartphones enables the convenient use of multimodal sensory substitution devices. Many applications such as EyeMusic [11] and Classic vOICe

---

[12] set the aforementioned algorithms to public access as means for large scale personal experiments. On the other hand, applications such as SeeingAI [13] or TapTapSee [14] provide users with verbal descriptions of captured images, making use of remote processing resources in a cloud computing schema.

III. MOTIVATION

The main challenge of sensory substitution devices is originating from a compromise made between the cognitive overload and the coded concepts of the input data. By concepts here the factors are meant determining the shape of the signal waveform with any level of dimensionality it takes. As an illustration, the intensity level and location of pixels in image data forms a meaningful pattern. Similarly, when investigating an audio stream made up of only pure sinusoids, the cardinality of the set, frequencies, and analogous amplitudes are operational in the resulting waveform. Considering the classic vOICe algorithm, there exists a direct mapping between the intensity levels and corresponding amplitudes. Henceforward, the vertical location of the pixel is related to the height of the pitch and the horizontal location to the latency of the occurring samples along the waveform.

The center of attention in the vOICe algorithm is the graphical pattern of the object represented in the gray scale image. Yet, other spatial concepts such as depth information and color are discarded. By channelizing the color information to the same scheme as performed in EyeMusic the cognitive load in recognizing the given pattern intensifies. The use of audio-tactile combinations might also be distracting or irritating by some records of skin reactions [15]. The frequency range of plausible sounds for the users, as well as their minimum distinguishable frequency interval, limits the size of the preimage in the visual domain [4] where not more than 60 to 64 distinguishable and pleasing sinusoid frequencies are available.

In the described mappings from visual to auditory domain the cognitive burden comes from the required attention and expertise in configuring the visual pattern based on the sound wave. Additionally, relating the configured pattern to a possible object in the environment based on the surrounding awareness is also a demanding task. In spite of the proficiency of the adept users in the experimental setting, the devices fail to preserve their applicability under real life circumstances due to the complexity of natural scenes.

The proposed method aims at reducing the processing load off the user's brain with means of a neural network responsible for the semantic segmentation of the given image.

IV. METHOD

Our method consists of several successive steps to convert a query input image into a comprehensible audio stream as illustrated in figure 2. The image is initially preprocessed by being resized into a 128×128 frame followed by a Convolutional Neural Network (CNN) responsible for background subtraction and object recognition. Then after, each detected object is assigned to a corresponding three-note audio permutation occurring in a specified time and pitch. Ultimately, the allocated sounds are combined to form a uniform audio stream. In the following, detailed explanation of each step is provided.

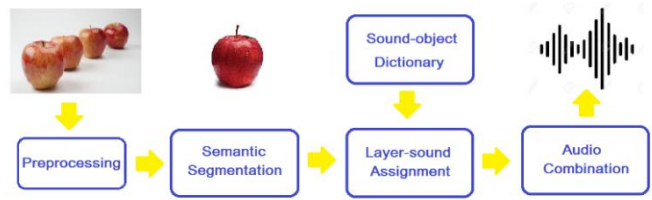

Fig 2. The proposed framework of the entire process

*A. Semantic segmentation and object recognition*

Detecting the meaningful objects in a rather complex environment can be a challenging task when no prior information about the background is available. The human visual perception however, puts seemingly miniscule effort on recognizing the region aligned with an object regardless of challenges such as occlusion and deformation effects [16] perhaps due to the fact of blob-based visual perception. The applied algorithm in recognizing the main entities within an image regardless of the background, relies on regional proposals which highly resembles the humanoid vision.

The proposed approach which was introduced in [17] efficiently detects the objects as well as simultaneous generation of a fine segmentation mask for each instance through a Recurrent Convolutional Neural Network (RCNN). The procedure is nominated as Mask-RCNN and is represented by its feedforward structure similar to the residual network of ResNet-101 [18] in figure 3.

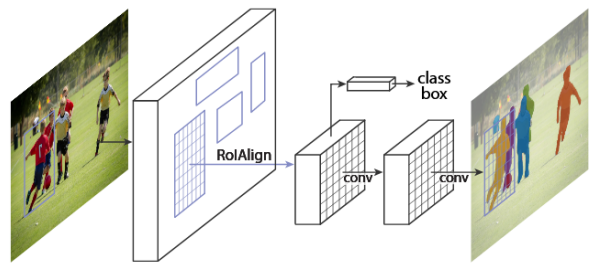

Fig. 3. The Mask-RCNN framework of instance segmentation [17].

Mask-RCNN is comprised of four distinct modules namely, feature extractors, Region Proposal Network (RPN), Region of Interest (ROI) classifier and bounding box regressor, and segmentation mask identifier. In the primary stage, a ResNet101 structure [18] is responsible for feature extraction hierarchically from the more generic visual patterns such as edges and corners to the more specific ones such as object appearances into a 32×32×2048 volume. The second stage is a shallow neural network to label the parts of the image as object-containing or not while scanning the image through a sliding window. By end of the third step, the class and bounding box refinement are available for each ROI. This network has deeper capacity in classifying the object present in the ROI as well as a fine tuning step regarding the boundary encapsulating the object. Lastly in the fourth module, the object-containing ROIs are fed to a neural network generating final masks corresponding to each noticeable object to identify its shape and location. Ultimately, the output of the network trained end-to-end on the COCO image dataset [19], is a volume with a face size similar to that of the image and a depth equal to the number of noticeable objects in it. Each layer of this volume overlays the mask of one entity.

In contrast with a classical RCNN segmenting network, each ROI in Mask-RCNN has a binary mask output independent from the prediction according to that region. This method uses bounding box classification and regression in parallel which is in nature simpler than RCNN itself. As shown in figure 3, the network head prediction of mask is separated from the classification network head resulting in the possibility of applying the multi-task loss function in equation (1).

$$L = L_{cls} + L_{box} + L_{mask} \quad (1)$$

This loss formulation includes the loss in classification of the object types defined on a per-pixel sigmoid, $L_{cls}$, localization, $L_{box}$, and segmentation mask as the average binary cross-entropy loss, $L_{mask}$. The weights of the network are tuned so that to minimize the loss function through an iterative Adam optimization algorithm as a combination of Stochastic Gradient Descent (SGD) and Root Mean Square Propagation (RMSprop) with a learning rate of 0.001 and momentum of 0.9. Figure 4 demonstrates a sample result of a query image into the semantic segmentation block.

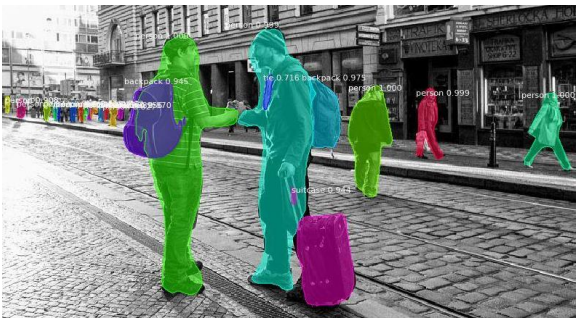

Fig. 4. Recognized objects by Mask-RCNN with their corresponding regions. The recognized class is labeled by Each ROI.

The resulted volume out of the Mask-RCNN contains a binary mask in each layer corresponding to an individual object in the image. The depth of the volume is determined by the number of detected objects in the image. In order to keep each of the objects tractable, the categories are limited to 86 normally seen objects which mostly occur as obstacles on the pathway of the visually impaired. For example, human-beings and cars are considered in the list whereas diamonds and curtains are discarded as rare or unimportant. Then after, each mask, primarily of the same size of the image, is quantized into an 8×8 grid. Each of the tiles in the quantized map makes it to be one if more than 10% of its region is covered by the original mask. Figure 5 illustrates the quantization procedure of a mask volume.

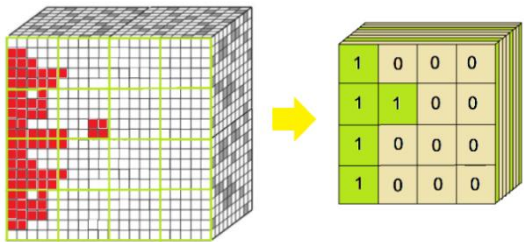

Fig. 5. Quantizing the binary Mask-RCNN labels into 8×8 grid.

### B. Object-sound dictionary

Given detected objects in each layer, we set up a dictionary assigning each object to particular three-note permutations. The elements of these sets are presented in table 1 as the notes of the lowest octave played by a flute.

Table 1. Basic notes for setting the dictionary words.

| Code | | Musical name | Frequency (Hz) |
|---|---|---|---|
| 1 | A | La | 55 |
| 2 | B | Si | 61.735 |
| 3 | C | Do | 65.406 |
| 4 | D | Re | 73.416 |
| 5 | E | Me | 82.407 |
| 6 | F | Fa | 87.307 |
| 7 | G | Sol | 97.999 |

Each dictionary word is a combination of the elements in table 1 related to one of the selected objects by the Mask-RCNN. The assigned combinations are given in table 2 for several common objects.

Table 2. Object to permutations assignment.

| object | Permutations (alphabetical) | Permutation (numerical) |
|---|---|---|
| person | AEA | 151 |
| chair | GCF | 736 |
| bed | GDA | 741 |
| dog | CAC | 313 |
| cat | AEG | 157 |
| banana | DDD | 444 |
| apple | GGB | 772 |
| cup | CGA | 227 |
| horse | GGG | 777 |
| book | ADC | 143 |

Figure 6 illustrates the given code words (sound permutations) in a notation similar to music script. Each row represents of the basic notes in table 1. The signs are read from left to right with the higher squares representing a higher frequency. This notation has proved useful in training process of the candidates for being intuitive and informative in perceiving the relative position of the notes both in an individual word and recognizing the words from one another.

The code words are produced based on seven general trends namely rising, strictly rising, falling, strictly falling, upward fluctuating, downward fluctuating, and constant. The bits of each code words stemming from each trend pattern must be selected so that to keep a threshold pairwise distance defined in equation (2). This, in turn, guarantees the detectability of the two words.

$$D = \sum_{i=1}^{number\ of\ bits} |a_i - b_i| \quad (2)$$

Where $a_i$ defines the bits in the first code word and $b_i$ the bits in the second.

### C. Layer-Sound Assignment

In the layer-sound assignment stage, every layer in the mask volume is matched with a one-second long audio stream. The voiced part in this waveform is generated in a specific pitch, length and latency in accordance with the object mask size and relative location. This section will provide points about how the audio stream is generated.

The higher tiles in a column attain higher pitches by octave intervals in between. For the aim of clarification, if and object mask covers a tile in the first row, the related code word will be made of basic notes given in table 1 and if the mask shields the i'th row, the code word will be made of notes given by equation (3).

$$Note_i = Note_1 \times 2^{i-1} \qquad (3)$$

Where $Note_i$ indicated the frequency of the element in the i'th row. If an object mask shades on multiple vertically adjacent, the summation of the code word in the corresponding octaves is played.

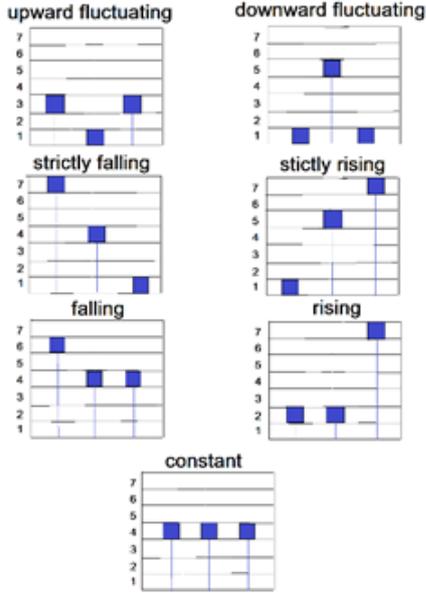

Fig. 6. Music-alike notation of the code words.

It is noteworthy that the highest note, G7, reaches 6271.936 Hz in frequency which is not obtained by an acoustic flute. Hence, synthetic sounds are applied. This high pitch was ensured to be audible for all the candidates. The intervals between the closest notes are defined surely larger than the minimum distinguishable interval by the acoustics of a flute. This margin increases as the notes grow higher in frequency. As mentioned earlier, the length of the stream is one second divided into eight successive parts retaining 1/8 second each. Wider objects correspond to longer voices parts in the audio stream. An n tile wide mask has a voiced part length of n/8 with n/(3×8) seconds taken by each note. In order to generate basic notes in various lengths in the time domain with the same pitch Time-Scale Modification (TSM) with a Hanning window is applied [20]. Figure 7 shows a two times stretched G1 note.

If two distinct objects of the same kind are detected in horizontally adjacent tiles, the related code will be played twice sequentially with length of each part according to one object. As a result, a wide object can be distinguished from a series of adjacent ones.

*D. Audio Combination*

Having generated the sounds allotted to single objects, the waveforms need to be combined to form the final frame audio. This is achieved through a simple linear scaling prior to addition of the waves so that to avoid saturation and overflows. Equation (4) describes the functionality of the audio combination block.

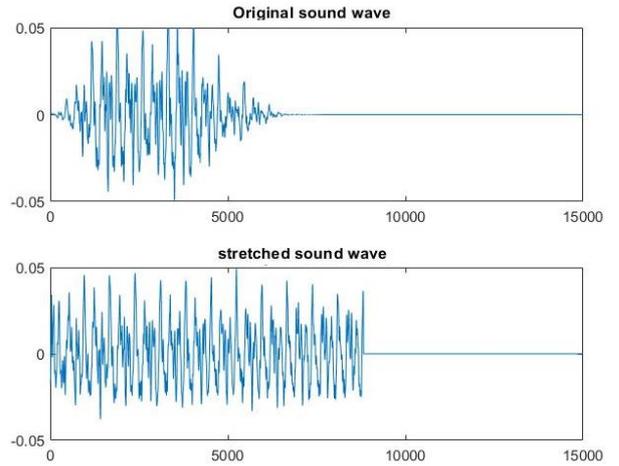

Fig. 7. Two times stretched $G_1$ in time domain using STFT.

$$S = \frac{1}{M} \sum_{i=1}^{M} S_i \qquad (4)$$

Where M is the number of detected objects in an image, Si is the audio stream generated for each of the objects, and S is the frame final audio.

V. EXPERIMENTAL RESULTS

A. *Training and Test Structure*

To assess the comprehensibility, plausibility, and pleasurableness of the sounds as well as the proficiency of the algorithm in delivering the visual information, a set of tests were designed. Each section of the training practice puts the main focus on one single aspect of information. In the light of detailed explanation, as the algorithm aims to transfer the object type, its size, and its relative location in the frame, each part is learned by the candidates separately. The primary stages of learning how to infer the sounds requires the user to recognize the object type regardless of its location and based on the three notes only. After mastering over the rise and falls in the pitch, the location of an individual object is variated along the two axis. The variations on the object size are introduced lastly where the applicant is asked to mark part of an empty 8×8 grid in which he/she supposes the object to be. Figure 8 shows a few sample images from the object recognition part.

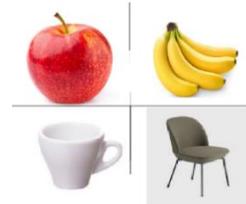

Fig. 8. Sample images for object recognition part.

Figure 9 illustrates an example of practices for locating an object with a fixed size over the grid. Figure 10 shows how inference of the object size is varied.

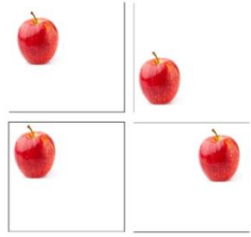

Fig. 9. Object location variations.

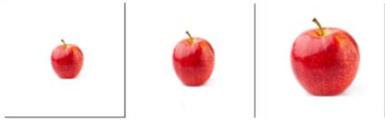

Fig. 10. Object size variations.

The final test comes in five sections object recognition; Test 1, Test 2 and Test 3: object recognition, Test 4: object coordinate, and Test 5: more than one object in the images which was not in the training part.

B. *Performance evaluation*

The average time spent on the training process for several candidates is 43.8 minutes on average which is noticeably shorter than the training time required to achieve the same rough level of comprehension through other sensory substitution devices which put the processing load on the user's brain. The results presented in table 3 come in five subsections explained in the test structure section. Yet it seems the number of applicants is not large enough and that they are highly influenced by their musical backgrounds (marked as MB in the table). The test design needs some elaboration on the third section to provide finer learning for applicants lacking a trained hearing.

Table 3. Test results.

| volunteer | | Test1 | Test2 | Test3 | Test4 | Test5 |
|---|---|---|---|---|---|---|
| age | MB | | | | | |
| 21 | No | 85.7% | 57.1% | 85.7% | 33.3% | 50% |
| 21 | Yes | 85.7% | 71.4% | 85.7% | 50% | 50% |
| 24 | Yes | 100% | 100% | 71.4% | 100% | 50% |
| 24 | No | 100% | 71.4% | 100% | 66.6% | 66.6% |
| 25 | No | 92.8% | 85.7% | 71.4% | 33.3% | 33.3% |
| 29 | No | 100% | 100% | 85.7% | 50% | 33.3% |
| 31 | Yes | 100% | 100% | 100% | 100% | 100% |

VI. CONCLUSION

This article introduced a novel algorithm for visual to auditory sensory substitution with the main goal of lowering the cognitive load on the user's brain. This is achieved by an object recognition and segmentation network in the image processing portion. The audio file is enhanced in terms of pleasurableness to a considerable extend by using synthetic musical notes rather than pure sinusoids. The increased pleasurableness is verified by the applicants who used the other similar versions such as the vOICe [5]. The training and testing packages yet need elaborations and extensions for blind users which are to be practiced in the future work.